\documentclass[aps,amsmath,amssymb,prl,superscriptaddress,longbibliography,twocolumn]{revtex4-1}

\usepackage{graphicx,color,dcolumn,bm,bbm,mathrsfs,amsfonts,amsmath,extarrows,mathtools,amssymb}
\DeclareMathAlphabet\mathbfcal{OMS}{cmsy}{b}{n}
\usepackage[colorlinks=true,citecolor=blue]{hyperref}
\hypersetup{colorlinks=true,citecolor=blue,linkcolor=red,urlcolor=blue}

\def\be{\begin{equation}}
\def\ee{\end{equation}}
\def\gammarr{\begin{array}{rll}}
\def\ea{\end{array}}
\def\bea{\begin{eqnarray}}
\def\eea{\end{eqnarray}}

\def\N2{$N{=}2$}

\def\>{\rangle}
\def\<{\langle}
\def\+{\dagger}
\def\={\ =\ }

\definecolor{grey}{rgb}{0.6, 0.6, 0.6}

\definecolor{purple}{rgb}{0.6, 0.0, 0.6}

\definecolor{green}{rgb}{0.0, 0.6, 0.0}

\begin{document}

\title{Synthetic gravitational horizons in low-dimensional quantum matter}

\author{Corentin Morice}
\thanks{These two authors contributed equally.}
\affiliation{Institute for Theoretical Physics and Delta Institute for Theoretical Physics, University of Amsterdam, 1090 GL Amsterdam, The Netherlands}
\author{Ali G. Moghaddam}
\thanks{These two authors contributed equally.}
\affiliation{Institute for Theoretical Solid State Physics, IFW Dresden, Helmholtzstr. 20, 01069 Dresden, Germany}
\address{Department of Physics, Institute for Advanced Studies in Basic Sciences (IASBS), Zanjan 45137-66731, Iran}
\author{Dmitry Chernyavsky}
\affiliation{Institute for Theoretical Solid State Physics, IFW Dresden, Helmholtzstr. 20, 01069 Dresden, Germany}
\author{Jasper van Wezel}
\affiliation{Institute for Theoretical Physics and Delta Institute for Theoretical Physics, University of Amsterdam, 1090 GL Amsterdam, The Netherlands}
\author{Jeroen van den Brink}
\affiliation{Institute for Theoretical Solid State Physics, IFW Dresden, Helmholtzstr. 20, 01069 Dresden, Germany}
\affiliation{Institute for Theoretical Physics and Würzburg-Dresden Cluster of Excellence ct.qmat, Technische Universität Dresden, 01069 Dresden, Germany}

\date{\today}

\begin{abstract}
We propose a class of lattice models realizable in a wide range of setups whose low-energy dynamics exactly reduces to Dirac fields subjected to (1+1)-dimensional gravitational backgrounds, including (anti-)de Sitter spacetime.
Wave-packets propagating on the lattice exhibit an eternal slowdown for power-law position-dependent hopping integrals $t(x)\propto x^\gamma$ when $\gamma\geq 1$, signalling the formation of black hole event horizons. 
For $\gamma< 1$ instead the wave-packets behave radically different and bounce off the horizon. 
We show that the eternal slowdown relates to a zero-energy spectral singularity of the lattice model and that the semiclassical wave packets trajectories coincide with the geodesics on (1+1)D dilaton gravity, paving the way for new and experimentally feasible routes to mimic black hole horizons and realize (1+1)D spacetimes as they appear in certain gravity theories.
\end{abstract}

\maketitle

\begin{figure}[t!]
\includegraphics[width=0.85\columnwidth]{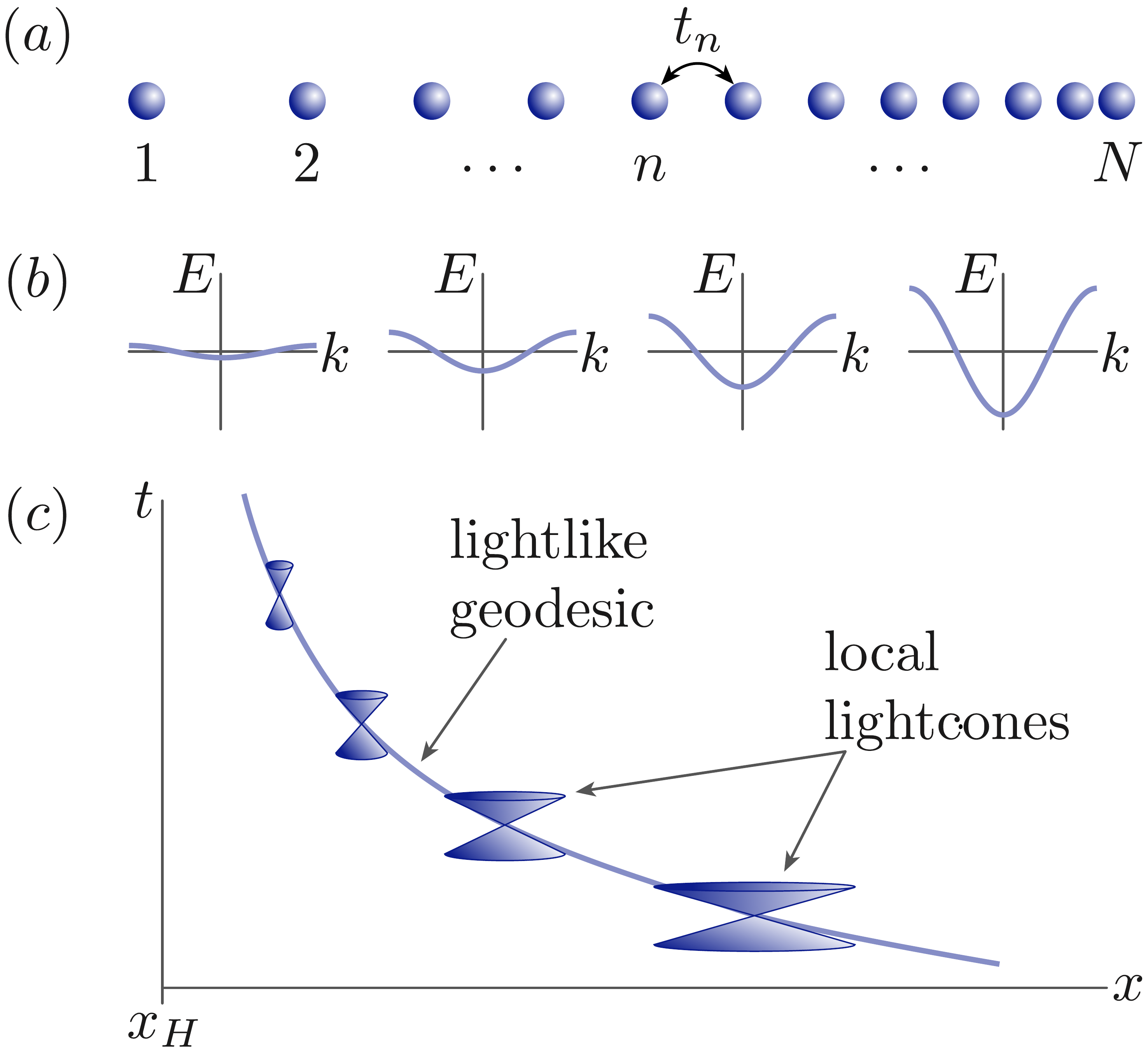}
\caption{{Sketch of the lattice-gravity correspondence} 
(a) A lattice model with position dependent hopping which mimics a curved spacetime and can possess event horizons.
(b) Schematics of the 2D black hole anti-de Sitter spacetime considered here. The velocity in spacetime diminishes at the vicinity of a static horizon $x_{H}$. The black curved line indicates a lightlike geodesic (worldline of free-falling massless particles) which exponentially approaches the horizons at $\tau \to \infty$. 
The local light cones attached to the worldlines are also shown at some points in spacetime. Moving from right to left the light cones always shrink, as dictated by $f_{\gamma=1}(x)=\alpha x$ in Eq. \eqref{eq:Dirac}.
}\label{fig0}
\end{figure}

Interesting subjects in physics may emerge from an original combination of ideas belonging to seemingly different areas of research. During the last decade for instance, theoretical methods and concepts from the realm of quantum gravity have opened up new areas in condensed matter physics. {\it Vice versa}, condensed matter systems can be created that closely resemble general relativity objects such as black holes \cite{volovik2003universe,Volovik:2016kid,novello2002artificial,hartnoll2011horizons,guan2017artificial,kedem2020black}, providing a playground to investigate certain aspects of black hole physics in an experimentally accessible setting.
In this context we consider the quantum evolution of wave-functions near an event horizon in the simplified situation of two spacetime dimensions, the lowest dimension exhibiting black holes. The advantage of considering (1+1)D spacetime is that we can make precise connections to wavefunction-dynamics of 1D quantum lattice systems with fermions or spins that can in principle be built and controlled in the lab.
In  particular, we construct 1D quantum lattice models whose low-energy properties reduce to the Dirac equation on a (1+1)D black hole background as it arises in Jackiw-Teitelboim (JT) gravity (see Fig. \ref{fig0}). In (1+1) spacetime dimensions, Einstein gravity is trivial and the JT theory represents the simplest gravity theory involving an additional scalar field called a dilaton. 
Within the condensed matter community, the JT gravity theory is known due to its relation to the Sachdev-Ye-Kitaev (SYK) model \cite{Sachdev1993,kitaev2015simple,kitaev2018,Maldacena2016prd,Maldacena2016}.

Here, we establish a direct relation between the geometry of the spacetime, particularly its geodesics, and the wave-packet dynamics in lattice models with hopping amplitudes that are explicit functions of the spatial coordinate. We explore the influence of the lattice, or equivalently the breaking of Lorentz invariance, which is seldom included in analogue gravity scenarios. The particles in the lattice model being essentially free allows for a combination of large-scale numerical simulations and semiclassical analytical approaches, the results of which we will show to concur.
In the low-energy limit, the quantum dynamics of the lattice models exactly follows that of a Dirac field subjected to a class of background metrics which appear as solutions of (1+1)D dilaton gravity. In particular, the case of linear position-dependence of the hopping parameter mimics (1+1)D anti-de Sitter space where particles exponentially slow down upon approaching a horizon. For general power-law position-dependence $t(x)\propto x^\gamma$, the local velocity vanishes at $x\to 0$ for any positive $\gamma$. For $0<\gamma<1$ the wave packets back scatter from $x=0$ and eternal slowdown takes place for $\gamma>1$. 
We show that the latter situation with slowdown of particles corresponds to the formation of a black hole horizon and interestingly is always associated with the presence of zero-energy spectral singularities of the lattice Hamiltonians. This observation bridges the analogue gravity features of the condensed matter system to its spectral singularity, remarkably an effect that went unnoticed in previous analogue gravity studies \cite{Barcelo2001,Cadoni2004,barcelo2011analogue,faccio2013analogue,Weststrom2017,huang2018black,Ojanen2019}.


\section*{2D dilaton gravity}
JT gravity is the simplest instance of a dilaton gravity in 2D.
Any solution of the JT gravity model with negative cosmological constant locally represents two-dimensional anti-de Sitter spacetime, which can be described by the metric \footnote{In order to bring this metric into the standard diagonalized form, one should redifine the temporal coordinate $\tau \rightarrow \tau+\frac{1}{2\alpha}\log(\alpha^2 x^2-1)$ }
\be\label{eq:metric_redefined}
ds^2 = -(\alpha^2 x^2-1)d\tau^2+2\alpha x \: dx \, d\tau-dx^2.
\ee
Globally this metric describes a (1+1)D black hole with the horizon points located at $x_{H^{\pm}}=\pm1/\alpha$, where the parameter $\alpha$ is related to the negative cosmological constant as $\alpha^2=-\Lambda$ (for a more detailed analysis of this black hole solution see \cite{Cadoni:1994uf}). 
Instead of concentrating only on the JT gravity, we consider a general dilaton theory with the metric of the form  \eqref{eq:metric_redefined} in which the linear function $\alpha x $ is replaced with a power-law form $f_{\gamma}(x)$ defined as $1 \pm f_\gamma(x) = (1\pm \alpha x )^\gamma$. The resulting metric, supported by the corresponding dilaton field, can be viewed as a solution of more general (1+1)D dilaton gravity theories \cite{Vassilevich2002}. 
The massless Dirac equation on this background takes the form
\begin{eqnarray}
\partial_\tau \Psi=\left(\sigma_3 \partial_x  - f_\gamma \sigma_0 \partial_x - \frac{1}{2}\frac{df_\gamma}{dx} \sigma_0 \right) \Psi
\label{eq:Dirac}
\end{eqnarray}
where $\Psi=\Psi(\tau,x)$ is a two-component spinor field and $\sigma_i$ with $i=0,\cdots,3$ denote the identity and Pauli matrices, respectively.
Eq.~\eqref{eq:Dirac} describes two linearly dispersing branches of fermions with a velocity that varies with the spatial coordinate via $f_\gamma(x)$. At $x_{H^{\pm}}$ the velocity of one of the two branches vanishes as $ (\alpha x)^\gamma$. We will show that for $\gamma \geq 1$ these two points represent an event horizon.


\section*{Lattice models with event horizon}
To connect this continuum field theory to condensed matter realizations, one needs to establish how it relates to the low-energy effective theory of a lattice Hamiltonian. The natural choice is to consider lattice models in which the band width depends on a spatial coordinate. For a continuum description to be valid, these variations should occur on length-scales much larger than the lattice constant. One can then picture the neighbourhood of any point in space as having constant hopping, and therefore a spectrum consisting in a single cosine band, where the group velocity at the Fermi wave vector and at half filling scales linearly with the band width.

Implementing this concept, we consider the Hamiltonian $\hat{H}$ on a 1D lattice of size $N$ with matrix elements
\begin{equation}
H_{nm} = -\left( \frac{n}{N-1} \right)^\gamma \delta_{n,m-1} - \left( \frac{m}{N-1} \right)^\gamma \delta_{n-1,m}
\label{eq:Hamiltonian_TB}
\end{equation}
which represent a coupling between neighboring sites $n$ and $m$, where $1\leq n\leq N-1$, see Fig. \ref{fig0}a.
The coupling strength between a site and its neighbor, also referred to as the hopping integral, is in the domain $(0,1]$ and depends on the position of the site as a powerlaw with exponent $\gamma$ .
The general Hamiltonian \eqref{eq:Hamiltonian_TB} is a particle-hole symmetric tight-binding (TB) model, which may be realized, for instance, in atomic chains placed by STM on a surface \cite{Eigler1990, Heinrich2006, Liljeroth2017}, ultracold atomic gasses in optical lattices \cite{zoller1998, lukin2003, bloch2008}, and photonic crystals \cite{joannopoulos1997, Ozawa2019}.
Apart from these direct realizations of position-dependent TB models, one may also consider other 1D settings which are equivalent to fermionic models on 1D lattices, such as the 1D XY model for spin-$1/2$, which maps to a fermionic problem via a Jordan-Wigner transformation \cite{mattis1961}.

\begin{figure}[t!]
\includegraphics[width=0.9\columnwidth]{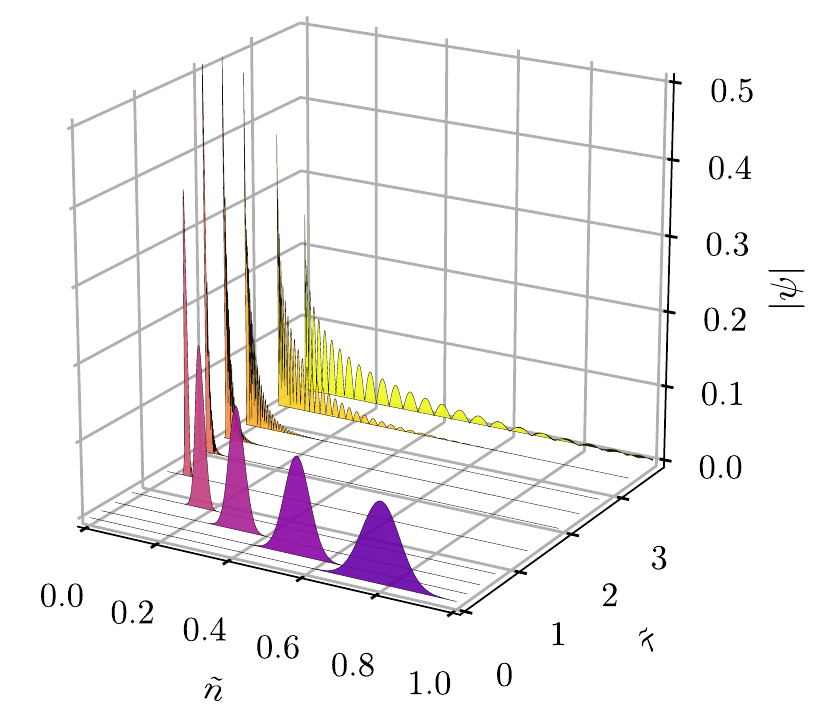}
\caption{\label{fig:wave-packet-evolution}
Time evolution of a Gaussian wave packet in the lattice model, with $\gamma=1$, $n_0=800$, $p_0=-\pi/2$, $w=50$, and $N=1001$. The wave packet slows down and localizes at the origin of the lattice, where it disintegrates.}
\end{figure}

The case $\gamma=0$ corresponds to a standard nearest-neighbour TB Hamiltonian with translation symmetry which, in the presence of periodic boundary conditions, gives rise to a band structure in reciprocal space. For $\gamma \neq 0$, translation symmetry is fundamentally broken and crystal momentum $k$ no longer is a good quantum number. Nevertheless, as outlined above and made more precise below, in the limit of large $N$, one can 
coarse-grain the full TB chain to smaller regions with 
locally constant hopping to which one can ascribe an approximate local band structure that reads $\varepsilon(n,k)\approx-2\,(n/N)^\gamma \cos k$
as a function of position $n$ and crystal momentum $k$.
Accordingly, at low energies ($|E| \ll 1$) which correspond to $k\sim \pm\pi/2$, the Hamiltonian \eqref{eq:Hamiltonian_TB} effectively describes a local Dirac dispersion with a position-dependent velocity resembling the Dirac field in JT background discussed above.
That way, the local band width $W(n)$ and local group velocity $v(n)\propto W(n)$ at the Fermi level will vary with the spatial coordinate as $n^\gamma$ similar to the hopping parameters. In this picture for $\gamma>0$, the group velocity vanishes when $n \to 0$, similarly to the group velocity of light cones approaching a black hole horizon, when viewed in asymptotic coordinates.
To determine how this intuitive picture translates into concrete physical phenomena,
we calculate the wave packet dynamics governed by Hamiltonian \eqref{eq:Hamiltonian_TB},
and ascertain under which circumstances wave packet propagation in the lattice models mimics geodesics in a black hole space time with a horizon at $n \to 0$.


\section*{Wave-packet dynamics \& semiclassical trajectories}
We first consider numerically the dynamics in the lattice model defined by Eq.\ \eqref{eq:Hamiltonian_TB} of a Gaussian wave packet with initial position $n_0\gg1$ and initial momentum $p_0$:
\begin{equation}
\psi(n,\tau=0) = \frac{1}{\sqrt[4]{\pi} \sqrt{w}} e^{-\frac{1}{2} \left( \frac{n-n_0}{w} \right) ^2} e^{i p_0 x},
\end{equation}
We first focus on $p_0 = -\pi/2$ which corresponds to a wave packet with an energy expectation value of zero.
The time evolution of the wave packet amplitude is represented in Fig. \ref{fig:wave-packet-evolution} for $\gamma=1$. The wave packet starts by sharpening while moving towards the origin of the lattice. As it comes very close to the origin, it starts forming ripples, which grow larger and larger, for $n/N$ larger than the position of the maximum. The wave packet eventually transforms entirely into incoherent ripples leaking out of the low $n$ region, while the maximum amplitude of the wave packet remains at the origin. 
The observed slowdown and localization of the wave packets towards $n/N \to 0$ indicate the presence of a horizon.
This feature emerges in the low-energy limit and remains intact as long as the width of the wave packet is large compared to the lattice spacing.
Then, after reaching a point of extreme localization, the wave packets start to disintegrate and form ripples which can be understood as a consequence of unitary evolution: two different wave packets cannot evolve into a single asymptotic limit, namely a Dirac delta-function distribution at the origin of the lattice.
Away from $p_0=-\pi/2$, i.e. when the wave packet has a finite energy expectation value, the point of the maximum amplitude of the wave packet starts propagating away from the origin well before reaching it.
In these cases the wave packet width decreases as it approaches the origin and again increases after being reflected.

\begin{figure}[t!]
\includegraphics[width=\columnwidth]{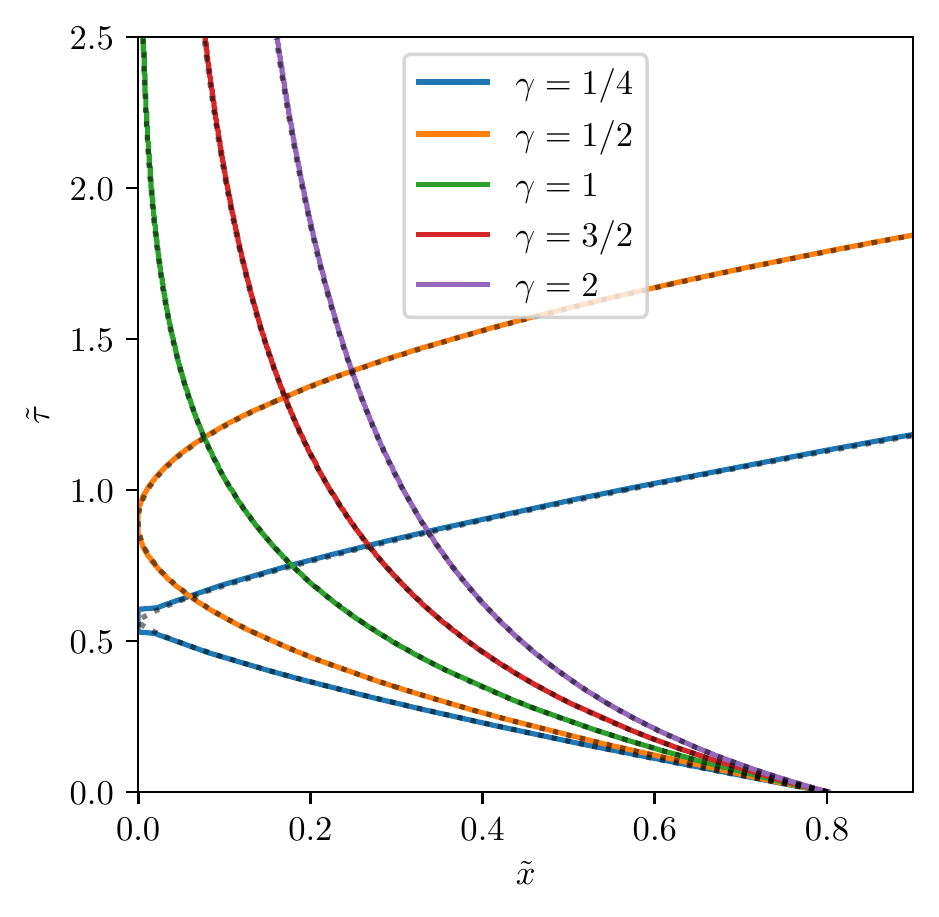}
\caption{\label{fig:wave-packet-evolution-exponent}
Position as a function of time of a Gaussian wave packet in the lattice model for five values of $\gamma$, with $x_0=800$, $p_0=-\pi/2$, $w=50$, and $N=1001$. The solid lines represent numerical calculations, while the dashed lines represent Eq.\ \eqref{eq:solution}. For $\gamma < 1$, the wave packet bounces on the origin of the lattice, while for $\gamma \geq 1$, it reaches the end of the lattice asymptotically.}
\end{figure}

We now turn to the influence of the form of the bandwidth's position-dependence, as controlled by $\gamma$, on the dynamics of wave packets with $p_0 = -\pi/2$ (Fig. \ref{fig:wave-packet-evolution-exponent}). We find that, for $ \gamma \geq 1$, wave packets slow down and asymptotically reach the origin of the lattice, which effectively behaves like a horizon for these wave packets. This asymptotic reaching of the origin is faster when decreasing $\gamma$, up to the point where it reaches 1. For $0 < \gamma < 1$, wave packets reach the origin and then start propagating away from it, thus effectively bouncing off the origin.
This observation suggests that only for $\gamma\geq1$, the lattice model with position-dependent hopping mimics the geodesics of particles falling into a black hole with horizon at $n=0$, as seen by a stationary distant observer.

To understand the evolution of the wave packet analytically, we derive their semiclassical trajectories. We define a continuous function $\psi(x)$ which, at each discrete lattice point, is equal to a solution of the lattice model $\psi_n$. We consider the Schr\"odinger equation for one row of the Hamiltonian $\hat{H}$ and assume that $\psi(x)$ is differentiable everywhere such that $\psi_{n \pm 1}$ can be replaced by its continuum expression $\psi(x \pm1)=e^{\pm i \hat{p}}\psi(x )$ in which $e^{\pm i \hat{p}}$ generates the finite translations $\Delta x=\pm 1$. From this one finds the effective Hamiltonian
\begin{eqnarray}
\tilde{\mathcal H}= 
- 
 \big(\frac{{\hat x}}{N-1}\big)^\gamma
 \: e^{ i \hat{p} }
 -e^{-i \hat{p} } \: \big(\frac{{\hat x}}{N-1}\big)^\gamma
 ,
\label{eq:effective-Hamiltonian}
\end{eqnarray}
which governs the dynamics of the wavefunction $\psi(x,t)$. Using this Hamiltonian, we derive the Heisenberg equations of motion for the momentum and position operators. We make a semiclassical approximation by setting the commutator of $\hat{x}$ and $\hat{p}$ to zero, which yields the equations
\begin{align}
&\frac{d{\tilde x}}{d{\tilde \tau}}
= 2 \, {\tilde x}^\gamma \, \sin {  p},
&\frac{d{ p}}{d{\tilde \tau}} = 2  \, \gamma {\tilde x}^{\gamma-1} \, \cos { p},
\label{eq:xdot-dynamics}
\end{align}
in which ${\tilde x}=x/(N-1)$ and ${\tilde \tau}=\tau/(N-1)$.
The above set of equations can be exactly solved in the general case using the hypergeometric function. Particularly for $p_0=-\pi/2$, the expectation value of the momentum operator remains constant in time and the expectation value of the position operator simplifies to
\begin{equation}
{\tilde x}({\tilde t}) = 
\begin{cases}
{\tilde x}_0 \, e^{-2{\tilde \tau}} & \text{if}~ \gamma=1
\\
\left| {\tilde x}_0^{1-\gamma}  -2 (1-\gamma)\, {\tilde \tau} \right|^{\frac{1}{1-\gamma}}  & \text{otherwise}
\end{cases}
\label{eq:solution}
\end{equation}
This time-evolution of the wave packet in the continuum agrees well with the exact time evolution using the lattice Hamiltonian defined in Eq. \eqref{eq:Hamiltonian_TB}, calculated numerically, as displayed in Fig. \ref{fig:wave-packet-evolution-exponent}. This is quite striking given the semiclassical nature of the continuum theory. The analytical trajectory in Eq.\ \eqref{eq:solution}, featuring the exponent $1/(1-\gamma)$, shines light on the fact that there is a transition at $\gamma=1$, between a regime where the wave packets localize at the origin of the lattice, which then behaves as a black hole horizon, and a regime where the wave packets bounce off the origin of the lattice.
The semiclassical analytical form now enables a direct comparison with geodesics and thus a relation with gravitational metrics.


\section*{Gravity/lattice-model equivalence}
Going back to the metric introduced in Eq.\ \eqref{eq:metric_redefined}, we consider the light-like geodesic equation $ds^2=0$. Assuming $f_{\gamma}=\pm \big(2x^\gamma-1\big)$, the light-like geodesics are given by $dx/d\tau=f_{\gamma}\pm1$ which are identical to Eq. \eqref{eq:xdot-dynamics} for $p_0=\pm\pi/2$, respectively. The trajectories of wave packets in the lattice model therefore coincide with light-like geodesics for the particles in a dilaton gravity background. Wave packets thus behave similarly in dilaton gravity and in the lattice model as long as the systems evolves within the bounds of validity of the semiclassical approach.

From the numerical simulations we see that wave packets on the lattice completely disintegrate in the long term after being extremely squeezed close to the horizon. This behaviour, which is absent in the case of a (continuum) quantum field subjected to a classical gravitational background, originates from the high energy cut-off that the lattice provides combined with unitary time evolution. Almost by definition an eternally slowing down wave packet close to the horizon cannot escape the discrete nature of the underlying lattice system -- short and long wavelength physics unavoidably couple so that the continuum description breaks down in finite time. 

\begin{figure}[t!]
\includegraphics[width=\columnwidth]{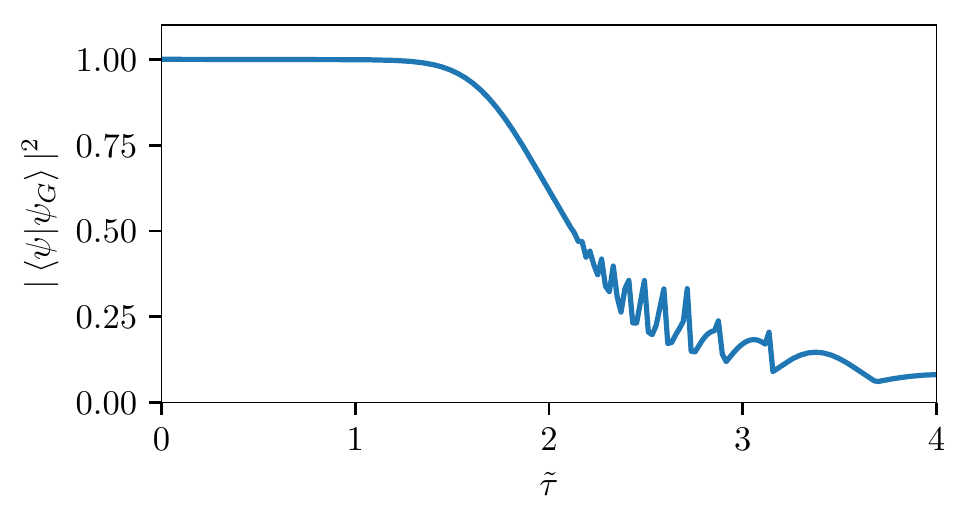}
\caption{\label{fig:wave-packet-overlap}
Overlap of $\psi_G(\tilde{\tau})$ and $\psi(\tilde{\tau})$ as a function of time. It stays almost constant up to the point where the wave packet reaches the proximity of the origin of the lattice.}
\end{figure}

By comparing to semiclassical trajectories one can further quantify this breakdown and the wave packet disintegration. Defining a Gaussian wave packet $\psi_G(\tilde{\tau})$ with position and width following Eq. \eqref{eq:solution} and a constant momentum $-\pi/2$, one can determine the overlap of $\psi_G(\tilde{\tau})$ with the time evolved original wave packet, see Fig.\ \ref{fig:wave-packet-overlap}. We observe that the overlap stays almost constant and equal to one up to the point in time where the wave packet gets very close to the origin and ripples start developing, at which point the overlap drops rapidly. The semiclassical description and its solution thus encode all the physics of the lattice system up to the point where the wave packet reaches the proximity of the horizon and breaks down when the wave packet width is of the order of the lattice spacing.

\section*{Relation to spectral properties}
\begin{figure}[t!]
\includegraphics[width=\columnwidth]{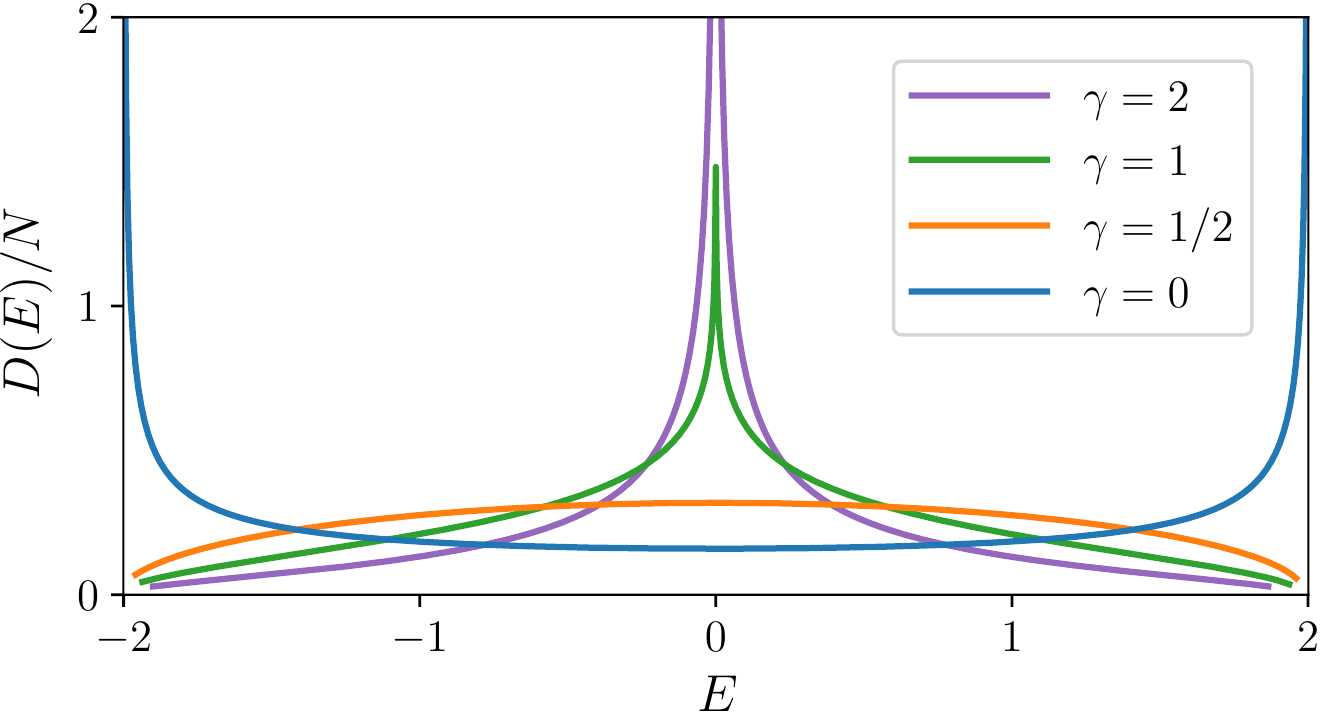}
\caption{\label{fig:DOS}
DOS $D(E)$ of the lattice model for four values of $\gamma$. While the uniform hopping model $t_n=1$ has van Hove singularities at the two ends of the band, for $\gamma\geq1$ a divergent DOS appears at $E=0$.
}
\end{figure}

We have shown in the above that $\gamma$ is a key parameter which governs a transition between two distinct behaviors via a critical regime where this approach is exponential for $\gamma = 1$. Strikingly, this transition is concomitant with a profound fingerprint in density of states (DOS), which exhibits a divergence at zero energy in the limit $N \rightarrow \infty$ for $\gamma \geq 1$ as shown in Fig. \ref{fig:DOS}. A further link between these two transitions which occur respectively in the behavior of particle dynamics at the vicinity of origin, and in the spectral properties of the lattice model, can be elucidated by the localized nature of eigenfunctions. Indeed, we observe that the states close to zero-energy always localize towards the origin. Therefore for $\gamma>1$, a very large DOS of the almost zero energy states all localized at the horizon, naturally accommodates the wave packets, whereas for $\gamma<1$, due the lack of a large DOS, the wave packets bounce off the origin in a finite time after being very close to it. This also sheds a light on the signification of $-\pi/2$ as a specific initial momentum of the wave packet which correspond to the zero-energy particles. Away from $E=0$, irrespective of the value of $\gamma$, we always have a finite spectral weight (DOS) and consequently the finite-energy wave packets always scatter off the origin. This highlights the role of zero energy, both in terms of wave packet energy and position in the spectrum, as the limit in which we find an equivalence between the evolution of wave packets in the lattice models and geodesics in gravitation theories.

\section*{Conclusions}
In the context of analogue gravity models a number of possible realizations have been proposed in different platforms such as acoustic and optical settings, cold atoms, and superconducting circuits \cite{Unruh:1980cg, Weinfurtner2011, philbin2008, steinhauer2016, Carusotto2008, kollar2019hyperbolic, Boettcher2020} which usually need precise tuning, and mimic the curved spacetime in an approximate way \cite{barcelo2011analogue}. Here we rather propose a direct relation between condensed matter systems and black hole metrics which are exact spacetime solutions of 2D dilaton gravity theories. 
We do so by considering one-dimensional lattice models realizable in solid-state, photonic and cold-atom settings to mimic the particle dynamics subjected to the two-dimensional gravity and particularly at the vicinity of horizons. Determining the dynamics of wave packets in lattice models with power-law position-dependence of the couplings $x^\gamma$, shows that the packets eternally slow down when $\gamma\geq1$, whereas they bounce back for $\gamma<1$. 
We have also found that the slowdown of wave packets, which signals the presence of a horizon, is concomitant with the divergence of the density of states at zero energy. The semiclassical wave packet trajectories coinciding with the geodesics on (1+1)D dilaton gravity provides a concrete and precise connection between the low-energy physics of the 1D lattice models and quantum fields subjected to (1+1)D gravity and opens a new perspective for condensed matter realizations of black holes and horizon physics.
In particular, it paves the way to explore quantum-mechanical aspects of black holes such as Hawking radiation and Unruh effect in an experimentally-accessible fashion.

\section*{Acknowledgements}

We thank Cosma Fulga, Flavio Nogueira and Lotte Mertens for stimulating discussions and acknowledge financial support through the Deutsche Forschungsgemeinschaft (DFG, German Research Foundation), through SFB 1143 project A5 and the Würzburg-Dresden Cluster of Excellence on Complexity and Topology in Quantum Matter- ct.qmat (EXC 2147, Project Id No. 390858490). A.G.M. acknowledges partial financial support from Iran Science Elites Federation under Grant No. 11/66332.

\bibliography{refs.bib}

\begin{thebibliography}{37}%
\makeatletter
\providecommand \@ifxundefined [1]{%
 \@ifx{#1\undefined}
}%
\providecommand \@ifnum [1]{%
 \ifnum #1\expandafter \@firstoftwo
 \else \expandafter \@secondoftwo
 \fi
}%
\providecommand \@ifx [1]{%
 \ifx #1\expandafter \@firstoftwo
 \else \expandafter \@secondoftwo
 \fi
}%
\providecommand \natexlab [1]{#1}%
\providecommand \enquote  [1]{``#1''}%
\providecommand \bibnamefont  [1]{#1}%
\providecommand \bibfnamefont [1]{#1}%
\providecommand \citenamefont [1]{#1}%
\providecommand \href@noop [0]{\@secondoftwo}%
\providecommand \href [0]{\begingroup \@sanitize@url \@href}%
\providecommand \@href[1]{\@@startlink{#1}\@@href}%
\providecommand \@@href[1]{\endgroup#1\@@endlink}%
\providecommand \@sanitize@url [0]{\catcode `\\12\catcode `\$12\catcode
  `\&12\catcode `\#12\catcode `\^12\catcode `\_12\catcode `\%12\relax}%
\providecommand \@@startlink[1]{}%
\providecommand \@@endlink[0]{}%
\providecommand \url  [0]{\begingroup\@sanitize@url \@url }%
\providecommand \@url [1]{\endgroup\@href {#1}{\urlprefix }}%
\providecommand \urlprefix  [0]{URL }%
\providecommand \Eprint [0]{\href }%
\providecommand \doibase [0]{http://dx.doi.org/}%
\providecommand \selectlanguage [0]{\@gobble}%
\providecommand \bibinfo  [0]{\@secondoftwo}%
\providecommand \bibfield  [0]{\@secondoftwo}%
\providecommand \translation [1]{[#1]}%
\providecommand \BibitemOpen [0]{}%
\providecommand \bibitemStop [0]{}%
\providecommand \bibitemNoStop [0]{.\EOS\space}%
\providecommand \EOS [0]{\spacefactor3000\relax}%
\providecommand \BibitemShut  [1]{\csname bibitem#1\endcsname}%
\let\auto@bib@innerbib\@empty
\bibitem [{\citenamefont {Volovik}(2003)}]{volovik2003universe}%
  \BibitemOpen
  \bibfield  {author} {\bibinfo {author} {\bibfnamefont {Grigory~E}\
  \bibnamefont {Volovik}},\ }\href@noop {} {\emph {\bibinfo {title} {The
  universe in a helium droplet}}},\ Vol.\ \bibinfo {volume} {117}\ (\bibinfo
  {publisher} {Oxford University Press, Oxford},\ \bibinfo {year}
  {2003})\BibitemShut {NoStop}%
\bibitem [{\citenamefont {Volovik}(2016)}]{Volovik:2016kid}%
  \BibitemOpen
  \bibfield  {author} {\bibinfo {author} {\bibfnamefont {Grigory~E.}\
  \bibnamefont {Volovik}},\ }\bibfield  {title} {\enquote {\bibinfo {title}
  {Black hole and hawking radiation by type-ii weyl fermions},}\ }\href
  {\doibase 10.1134/S0021364016210050} {\bibfield  {journal} {\bibinfo
  {journal} {JETP Lett.}\ }\textbf {\bibinfo {volume} {104}},\ \bibinfo {pages}
  {645} (\bibinfo {year} {2016})}\BibitemShut {NoStop}%
\bibitem [{\citenamefont {Novello}\ \emph {et~al.}(2002)\citenamefont
  {Novello}, \citenamefont {Visser},\ and\ \citenamefont
  {Volovik}}]{novello2002artificial}%
  \BibitemOpen
  \bibfield  {author} {\bibinfo {author} {\bibfnamefont {M{\'a}rio}\
  \bibnamefont {Novello}}, \bibinfo {author} {\bibfnamefont {Matt}\
  \bibnamefont {Visser}}, \ and\ \bibinfo {author} {\bibfnamefont {Grigory~E}\
  \bibnamefont {Volovik}},\ }\href@noop {} {\emph {\bibinfo {title} {Artificial
  black holes}}}\ (\bibinfo  {publisher} {World Scientific},\ \bibinfo {year}
  {2002})\BibitemShut {NoStop}%
\bibitem [{\citenamefont {Hartnoll}(2011)}]{hartnoll2011horizons}%
  \BibitemOpen
  \bibfield  {author} {\bibinfo {author} {\bibfnamefont {Sean~A.}\ \bibnamefont
  {Hartnoll}},\ }\href@noop {} {\enquote {\bibinfo {title} {Horizons,
  holography and condensed matter},}\ } (\bibinfo {year} {2011}),\ \Eprint
  {http://arxiv.org/abs/1106.4324} {arXiv:1106.4324 [hep-th]} \BibitemShut
  {NoStop}%
\bibitem [{\citenamefont {Guan}\ \emph {et~al.}(2017)\citenamefont {Guan},
  \citenamefont {Yu}, \citenamefont {Liu}, \citenamefont {Liu}, \citenamefont
  {Dong}, \citenamefont {Lu}, \citenamefont {Yao},\ and\ \citenamefont
  {Yang}}]{guan2017artificial}%
  \BibitemOpen
  \bibfield  {author} {\bibinfo {author} {\bibfnamefont {Shan}\ \bibnamefont
  {Guan}}, \bibinfo {author} {\bibfnamefont {Zhi-Ming}\ \bibnamefont {Yu}},
  \bibinfo {author} {\bibfnamefont {Ying}\ \bibnamefont {Liu}}, \bibinfo
  {author} {\bibfnamefont {Gui-Bin}\ \bibnamefont {Liu}}, \bibinfo {author}
  {\bibfnamefont {Liang}\ \bibnamefont {Dong}}, \bibinfo {author}
  {\bibfnamefont {Yunhao}\ \bibnamefont {Lu}}, \bibinfo {author} {\bibfnamefont
  {Yugui}\ \bibnamefont {Yao}}, \ and\ \bibinfo {author} {\bibfnamefont
  {Shengyuan~A}\ \bibnamefont {Yang}},\ }\bibfield  {title} {\enquote {\bibinfo
  {title} {Artificial gravity field, astrophysical analogues, and topological
  phase transitions in strained topological semimetals},}\ }\href {\doibase
  10.1038/s41535-017-0026-7} {\bibfield  {journal} {\bibinfo  {journal} {npj
  Quantum Mater.}\ }\textbf {\bibinfo {volume} {2}},\ \bibinfo {pages} {1}
  (\bibinfo {year} {2017})}\BibitemShut {NoStop}%
\bibitem [{\citenamefont {Kedem}\ \emph {et~al.}(2020)\citenamefont {Kedem},
  \citenamefont {Bergholtz},\ and\ \citenamefont {Wilczek}}]{kedem2020black}%
  \BibitemOpen
  \bibfield  {author} {\bibinfo {author} {\bibfnamefont {Yaron}\ \bibnamefont
  {Kedem}}, \bibinfo {author} {\bibfnamefont {Emil~J.}\ \bibnamefont
  {Bergholtz}}, \ and\ \bibinfo {author} {\bibfnamefont {Frank}\ \bibnamefont
  {Wilczek}},\ }\bibfield  {title} {\enquote {\bibinfo {title} {Black and white
  holes at material junctions},}\ }\href {\doibase
  10.1103/PhysRevResearch.2.043285} {\bibfield  {journal} {\bibinfo  {journal}
  {Phys. Rev. Research}\ }\textbf {\bibinfo {volume} {2}},\ \bibinfo {pages}
  {043285} (\bibinfo {year} {2020})}\BibitemShut {NoStop}%
\bibitem [{\citenamefont {Sachdev}\ and\ \citenamefont
  {Ye}(1993)}]{Sachdev1993}%
  \BibitemOpen
  \bibfield  {author} {\bibinfo {author} {\bibfnamefont {Subir}\ \bibnamefont
  {Sachdev}}\ and\ \bibinfo {author} {\bibfnamefont {Jinwu}\ \bibnamefont
  {Ye}},\ }\bibfield  {title} {\enquote {\bibinfo {title} {Gapless spin-fluid
  ground state in a random quantum heisenberg magnet},}\ }\href {\doibase
  10.1103/PhysRevLett.70.3339} {\bibfield  {journal} {\bibinfo  {journal}
  {Phys. Rev. Lett.}\ }\textbf {\bibinfo {volume} {70}},\ \bibinfo {pages}
  {3339} (\bibinfo {year} {1993})}\BibitemShut {NoStop}%
\bibitem [{\citenamefont {Kitaev}(2015)}]{kitaev2015simple}%
  \BibitemOpen
  \bibfield  {author} {\bibinfo {author} {\bibfnamefont {Alexei}\ \bibnamefont
  {Kitaev}},\ }\bibfield  {title} {\enquote {\bibinfo {title} {A simple model
  of quantum holography},}\ }in\ \href@noop {} {\emph {\bibinfo {booktitle}
  {KITP strings seminar and Entanglement}}},\ Vol.~\bibinfo {volume} {12}\
  (\bibinfo {year} {2015})\ p.~\bibinfo {pages} {26}\BibitemShut {NoStop}%
\bibitem [{\citenamefont {Kitaev}\ and\ \citenamefont
  {Suh}(2018)}]{kitaev2018}%
  \BibitemOpen
  \bibfield  {author} {\bibinfo {author} {\bibfnamefont {Alexei}\ \bibnamefont
  {Kitaev}}\ and\ \bibinfo {author} {\bibfnamefont {S~Josephine}\ \bibnamefont
  {Suh}},\ }\bibfield  {title} {\enquote {\bibinfo {title} {The soft mode in
  the sachdev-ye-kitaev model and its gravity dual},}\ }\href {\doibase
  10.1007/JHEP05(2018)183} {\bibfield  {journal} {\bibinfo  {journal} {J. High
  Energy Phys.}\ }\textbf {\bibinfo {volume} {2018}},\ \bibinfo {pages} {183}
  (\bibinfo {year} {2018})}\BibitemShut {NoStop}%
\bibitem [{\citenamefont {Maldacena}\ and\ \citenamefont
  {Stanford}(2016)}]{Maldacena2016prd}%
  \BibitemOpen
  \bibfield  {author} {\bibinfo {author} {\bibfnamefont {Juan}\ \bibnamefont
  {Maldacena}}\ and\ \bibinfo {author} {\bibfnamefont {Douglas}\ \bibnamefont
  {Stanford}},\ }\bibfield  {title} {\enquote {\bibinfo {title} {Remarks on the
  sachdev-ye-kitaev model},}\ }\href {\doibase 10.1103/PhysRevD.94.106002}
  {\bibfield  {journal} {\bibinfo  {journal} {Phys. Rev. D}\ }\textbf {\bibinfo
  {volume} {94}},\ \bibinfo {pages} {106002} (\bibinfo {year}
  {2016})}\BibitemShut {NoStop}%
\bibitem [{\citenamefont {Maldacena}\ \emph {et~al.}(2016)\citenamefont
  {Maldacena}, \citenamefont {Stanford},\ and\ \citenamefont
  {Yang}}]{Maldacena2016}%
  \BibitemOpen
  \bibfield  {author} {\bibinfo {author} {\bibfnamefont {Juan}\ \bibnamefont
  {Maldacena}}, \bibinfo {author} {\bibfnamefont {Douglas}\ \bibnamefont
  {Stanford}}, \ and\ \bibinfo {author} {\bibfnamefont {Zhenbin}\ \bibnamefont
  {Yang}},\ }\bibfield  {title} {\enquote {\bibinfo {title} {Conformal symmetry
  and its breaking in two-dimensional nearly anti-de sitter space},}\ }\href
  {\doibase 10.1093/ptep/ptw124} {\bibfield  {journal} {\bibinfo  {journal}
  {Prog. Theo. Exp. Phys.}\ }\textbf {\bibinfo {volume} {2016}},\ \bibinfo
  {pages} {12C104} (\bibinfo {year} {2016})}\BibitemShut {NoStop}%
\bibitem [{\citenamefont {Barcel{\'{o}}}\ \emph {et~al.}(2001)\citenamefont
  {Barcel{\'{o}}}, \citenamefont {Liberati},\ and\ \citenamefont
  {Visser}}]{Barcelo2001}%
  \BibitemOpen
  \bibfield  {author} {\bibinfo {author} {\bibfnamefont {Carlos}\ \bibnamefont
  {Barcel{\'{o}}}}, \bibinfo {author} {\bibfnamefont {S}~\bibnamefont
  {Liberati}}, \ and\ \bibinfo {author} {\bibfnamefont {Matt}\ \bibnamefont
  {Visser}},\ }\bibfield  {title} {\enquote {\bibinfo {title} {Analogue gravity
  from bose-einstein condensates},}\ }\href {\doibase
  10.1088/0264-9381/18/6/312} {\bibfield  {journal} {\bibinfo  {journal}
  {Class. Quantum Gravity}\ }\textbf {\bibinfo {volume} {18}},\ \bibinfo
  {pages} {1137} (\bibinfo {year} {2001})}\BibitemShut {NoStop}%
\bibitem [{\citenamefont {Cadoni}(2004)}]{Cadoni2004}%
  \BibitemOpen
  \bibfield  {author} {\bibinfo {author} {\bibfnamefont {Mariano}\ \bibnamefont
  {Cadoni}},\ }\bibfield  {title} {\enquote {\bibinfo {title} {Acoustic
  analogues of two-dimensional black holes},}\ }\href {\doibase
  10.1088/0264-9381/22/2/012} {\bibfield  {journal} {\bibinfo  {journal}
  {Classical and Quantum Gravity}\ }\textbf {\bibinfo {volume} {22}},\ \bibinfo
  {pages} {409} (\bibinfo {year} {2004})}\BibitemShut {NoStop}%
\bibitem [{\citenamefont {Barcel{\'o}}\ \emph {et~al.}(2011)\citenamefont
  {Barcel{\'o}}, \citenamefont {Liberati},\ and\ \citenamefont
  {Visser}}]{barcelo2011analogue}%
  \BibitemOpen
  \bibfield  {author} {\bibinfo {author} {\bibfnamefont {Carlos}\ \bibnamefont
  {Barcel{\'o}}}, \bibinfo {author} {\bibfnamefont {Stefano}\ \bibnamefont
  {Liberati}}, \ and\ \bibinfo {author} {\bibfnamefont {Matt}\ \bibnamefont
  {Visser}},\ }\bibfield  {title} {\enquote {\bibinfo {title} {Analogue
  gravity},}\ }\href {\doibase 10.12942/lrr-2011-3} {\bibfield  {journal}
  {\bibinfo  {journal} {Living Rev. Relativ.}\ }\textbf {\bibinfo {volume}
  {14}},\ \bibinfo {pages} {3} (\bibinfo {year} {2011})}\BibitemShut {NoStop}%
\bibitem [{\citenamefont {Faccio}\ \emph {et~al.}(2013)\citenamefont {Faccio},
  \citenamefont {Belgiorno}, \citenamefont {Cacciatori}, \citenamefont
  {Gorini}, \citenamefont {Liberati},\ and\ \citenamefont
  {Moschella}}]{faccio2013analogue}%
  \BibitemOpen
  \bibinfo {editor} {\bibfnamefont {Daniele}\ \bibnamefont {Faccio}}, \bibinfo
  {editor} {\bibfnamefont {Francesco}\ \bibnamefont {Belgiorno}}, \bibinfo
  {editor} {\bibfnamefont {Sergio}\ \bibnamefont {Cacciatori}}, \bibinfo
  {editor} {\bibfnamefont {Vittorio}\ \bibnamefont {Gorini}}, \bibinfo {editor}
  {\bibfnamefont {Stefano}\ \bibnamefont {Liberati}}, \ and\ \bibinfo {editor}
  {\bibfnamefont {Ugo}\ \bibnamefont {Moschella}},\ eds.,\ \href {\doibase
  10.1007/978-3-319-00266-8} {\emph {\bibinfo {title} {Analogue gravity
  phenomenology: analogue spacetimes and horizons, from theory to
  experiment}}},\ \bibinfo {series} {Lecture Notes in Physics}, Vol.\ \bibinfo
  {volume} {870}\ (\bibinfo  {publisher} {Springer},\ \bibinfo {address}
  {Cham},\ \bibinfo {year} {2013})\BibitemShut {NoStop}%
\bibitem [{\citenamefont {Weststr\"om}\ and\ \citenamefont
  {Ojanen}(2017)}]{Weststrom2017}%
  \BibitemOpen
  \bibfield  {author} {\bibinfo {author} {\bibfnamefont {Alex}\ \bibnamefont
  {Weststr\"om}}\ and\ \bibinfo {author} {\bibfnamefont {Teemu}\ \bibnamefont
  {Ojanen}},\ }\bibfield  {title} {\enquote {\bibinfo {title} {Designer
  curved-space geometry for relativistic fermions in weyl metamaterials},}\
  }\href {\doibase 10.1103/PhysRevX.7.041026} {\bibfield  {journal} {\bibinfo
  {journal} {Phys. Rev. X}\ }\textbf {\bibinfo {volume} {7}},\ \bibinfo {pages}
  {041026} (\bibinfo {year} {2017})}\BibitemShut {NoStop}%
\bibitem [{\citenamefont {Huang}\ \emph {et~al.}(2018)\citenamefont {Huang},
  \citenamefont {Jin},\ and\ \citenamefont {Liu}}]{huang2018black}%
  \BibitemOpen
  \bibfield  {author} {\bibinfo {author} {\bibfnamefont {Huaqing}\ \bibnamefont
  {Huang}}, \bibinfo {author} {\bibfnamefont {Kyung-Hwan}\ \bibnamefont {Jin}},
  \ and\ \bibinfo {author} {\bibfnamefont {Feng}\ \bibnamefont {Liu}},\
  }\bibfield  {title} {\enquote {\bibinfo {title} {Black-hole horizon in the
  dirac semimetal ${\mathrm{zn}}_{2}{\mathrm{in}}_{2}{\mathrm{s}}_{5}$},}\
  }\href {\doibase 10.1103/PhysRevB.98.121110} {\bibfield  {journal} {\bibinfo
  {journal} {Phys. Rev. B}\ }\textbf {\bibinfo {volume} {98}},\ \bibinfo
  {pages} {121110} (\bibinfo {year} {2018})}\BibitemShut {NoStop}%
\bibitem [{\citenamefont {Liang}\ and\ \citenamefont
  {Ojanen}(2019)}]{Ojanen2019}%
  \BibitemOpen
  \bibfield  {author} {\bibinfo {author} {\bibfnamefont {Long}\ \bibnamefont
  {Liang}}\ and\ \bibinfo {author} {\bibfnamefont {Teemu}\ \bibnamefont
  {Ojanen}},\ }\bibfield  {title} {\enquote {\bibinfo {title} {Curved spacetime
  theory of inhomogeneous weyl materials},}\ }\href {\doibase
  10.1103/PhysRevResearch.1.032006} {\bibfield  {journal} {\bibinfo  {journal}
  {Phys. Rev. Research}\ }\textbf {\bibinfo {volume} {1}},\ \bibinfo {pages}
  {032006} (\bibinfo {year} {2019})}\BibitemShut {NoStop}%
\bibitem [{Note1()}]{Note1}%
  \BibitemOpen
  \bibinfo {note} {In order to bring this metric into the standard diagonalized
  form, one should redifine the temporal coordinate $\tau \rightarrow \tau
  +\protect \frac {1}{2\alpha }\protect \qopname \relax o{log}(\alpha ^2
  x^2-1)$}\BibitemShut {NoStop}%
\bibitem [{\citenamefont {Cadoni}\ and\ \citenamefont
  {Mignemi}(1995)}]{Cadoni:1994uf}%
  \BibitemOpen
  \bibfield  {author} {\bibinfo {author} {\bibfnamefont {Mariano}\ \bibnamefont
  {Cadoni}}\ and\ \bibinfo {author} {\bibfnamefont {Salvatore}\ \bibnamefont
  {Mignemi}},\ }\bibfield  {title} {\enquote {\bibinfo {title} {{Nonsingular
  four-dimensional black holes and the Jackiw-Teitelboim theory}},}\ }\href
  {\doibase 10.1103/PhysRevD.51.4319} {\bibfield  {journal} {\bibinfo
  {journal} {Phys. Rev. D}\ }\textbf {\bibinfo {volume} {51}},\ \bibinfo
  {pages} {4319} (\bibinfo {year} {1995})}\BibitemShut {NoStop}%
\bibitem [{\citenamefont {Grumiller}\ \emph {et~al.}(2002)\citenamefont
  {Grumiller}, \citenamefont {Kummer},\ and\ \citenamefont
  {Vassilevich}}]{Vassilevich2002}%
  \BibitemOpen
  \bibfield  {author} {\bibinfo {author} {\bibfnamefont {D.}~\bibnamefont
  {Grumiller}}, \bibinfo {author} {\bibfnamefont {W.}~\bibnamefont {Kummer}}, \
  and\ \bibinfo {author} {\bibfnamefont {D.V.}\ \bibnamefont {Vassilevich}},\
  }\bibfield  {title} {\enquote {\bibinfo {title} {Dilaton gravity in two
  dimensions},}\ }\href {\doibase
  https://doi.org/10.1016/S0370-1573(02)00267-3} {\bibfield  {journal}
  {\bibinfo  {journal} {Phys. Rep.}\ }\textbf {\bibinfo {volume} {369}},\
  \bibinfo {pages} {327} (\bibinfo {year} {2002})}\BibitemShut {NoStop}%
\bibitem [{\citenamefont {Eigler}\ and\ \citenamefont
  {Schweizer}(1990)}]{Eigler1990}%
  \BibitemOpen
  \bibfield  {author} {\bibinfo {author} {\bibfnamefont {Donald~M}\
  \bibnamefont {Eigler}}\ and\ \bibinfo {author} {\bibfnamefont {Erhard~K}\
  \bibnamefont {Schweizer}},\ }\bibfield  {title} {\enquote {\bibinfo {title}
  {Positioning single atoms with a scanning tunnelling microscope},}\ }\href
  {\doibase 10.1038/344524a0} {\bibfield  {journal} {\bibinfo  {journal}
  {Nature}\ }\textbf {\bibinfo {volume} {344}},\ \bibinfo {pages} {524}
  (\bibinfo {year} {1990})}\BibitemShut {NoStop}%
\bibitem [{\citenamefont {Hirjibehedin}\ \emph {et~al.}(2006)\citenamefont
  {Hirjibehedin}, \citenamefont {Lutz},\ and\ \citenamefont
  {Heinrich}}]{Heinrich2006}%
  \BibitemOpen
  \bibfield  {author} {\bibinfo {author} {\bibfnamefont {Cyrus~F}\ \bibnamefont
  {Hirjibehedin}}, \bibinfo {author} {\bibfnamefont {Christopher~P}\
  \bibnamefont {Lutz}}, \ and\ \bibinfo {author} {\bibfnamefont {Andreas~J}\
  \bibnamefont {Heinrich}},\ }\bibfield  {title} {\enquote {\bibinfo {title}
  {Spin coupling in engineered atomic structures},}\ }\href {\doibase
  10.1126/science.1125398} {\bibfield  {journal} {\bibinfo  {journal}
  {Science}\ }\textbf {\bibinfo {volume} {312}},\ \bibinfo {pages} {1021}
  (\bibinfo {year} {2006})}\BibitemShut {NoStop}%
\bibitem [{\citenamefont {Drost}\ \emph {et~al.}(2017)\citenamefont {Drost},
  \citenamefont {Ojanen}, \citenamefont {Harju},\ and\ \citenamefont
  {Liljeroth}}]{Liljeroth2017}%
  \BibitemOpen
  \bibfield  {author} {\bibinfo {author} {\bibfnamefont {Robert}\ \bibnamefont
  {Drost}}, \bibinfo {author} {\bibfnamefont {Teemu}\ \bibnamefont {Ojanen}},
  \bibinfo {author} {\bibfnamefont {Ari}\ \bibnamefont {Harju}}, \ and\
  \bibinfo {author} {\bibfnamefont {Peter}\ \bibnamefont {Liljeroth}},\
  }\bibfield  {title} {\enquote {\bibinfo {title} {Topological states in
  engineered atomic lattices},}\ }\href {\doibase 10.1038/nphys4080} {\bibfield
   {journal} {\bibinfo  {journal} {Nat. Phys.}\ }\textbf {\bibinfo {volume}
  {13}},\ \bibinfo {pages} {668} (\bibinfo {year} {2017})}\BibitemShut
  {NoStop}%
\bibitem [{\citenamefont {Jaksch}\ \emph {et~al.}(1998)\citenamefont {Jaksch},
  \citenamefont {Bruder}, \citenamefont {Cirac}, \citenamefont {Gardiner},\
  and\ \citenamefont {Zoller}}]{zoller1998}%
  \BibitemOpen
  \bibfield  {author} {\bibinfo {author} {\bibfnamefont {D.}~\bibnamefont
  {Jaksch}}, \bibinfo {author} {\bibfnamefont {C.}~\bibnamefont {Bruder}},
  \bibinfo {author} {\bibfnamefont {J.~I.}\ \bibnamefont {Cirac}}, \bibinfo
  {author} {\bibfnamefont {C.~W.}\ \bibnamefont {Gardiner}}, \ and\ \bibinfo
  {author} {\bibfnamefont {P.}~\bibnamefont {Zoller}},\ }\bibfield  {title}
  {\enquote {\bibinfo {title} {Cold bosonic atoms in optical lattices},}\
  }\href {\doibase 10.1103/PhysRevLett.81.3108} {\bibfield  {journal} {\bibinfo
   {journal} {Phys. Rev. Lett.}\ }\textbf {\bibinfo {volume} {81}},\ \bibinfo
  {pages} {3108} (\bibinfo {year} {1998})}\BibitemShut {NoStop}%
\bibitem [{\citenamefont {Duan}\ \emph {et~al.}(2003)\citenamefont {Duan},
  \citenamefont {Demler},\ and\ \citenamefont {Lukin}}]{lukin2003}%
  \BibitemOpen
  \bibfield  {author} {\bibinfo {author} {\bibfnamefont {L.-M.}\ \bibnamefont
  {Duan}}, \bibinfo {author} {\bibfnamefont {E.}~\bibnamefont {Demler}}, \ and\
  \bibinfo {author} {\bibfnamefont {M.~D.}\ \bibnamefont {Lukin}},\ }\bibfield
  {title} {\enquote {\bibinfo {title} {Controlling spin exchange interactions
  of ultracold atoms in optical lattices},}\ }\href {\doibase
  10.1103/PhysRevLett.91.090402} {\bibfield  {journal} {\bibinfo  {journal}
  {Phys. Rev. Lett.}\ }\textbf {\bibinfo {volume} {91}},\ \bibinfo {pages}
  {090402} (\bibinfo {year} {2003})}\BibitemShut {NoStop}%
\bibitem [{\citenamefont {Bloch}\ \emph {et~al.}(2008)\citenamefont {Bloch},
  \citenamefont {Dalibard},\ and\ \citenamefont {Zwerger}}]{bloch2008}%
  \BibitemOpen
  \bibfield  {author} {\bibinfo {author} {\bibfnamefont {Immanuel}\
  \bibnamefont {Bloch}}, \bibinfo {author} {\bibfnamefont {Jean}\ \bibnamefont
  {Dalibard}}, \ and\ \bibinfo {author} {\bibfnamefont {Wilhelm}\ \bibnamefont
  {Zwerger}},\ }\bibfield  {title} {\enquote {\bibinfo {title} {Many-body
  physics with ultracold gases},}\ }\href {\doibase 10.1103/RevModPhys.80.885}
  {\bibfield  {journal} {\bibinfo  {journal} {Rev. Mod. Phys.}\ }\textbf
  {\bibinfo {volume} {80}},\ \bibinfo {pages} {885} (\bibinfo {year}
  {2008})}\BibitemShut {NoStop}%
\bibitem [{\citenamefont {Joannopoulos}\ \emph {et~al.}(1997)\citenamefont
  {Joannopoulos}, \citenamefont {Villeneuve},\ and\ \citenamefont
  {Fan}}]{joannopoulos1997}%
  \BibitemOpen
  \bibfield  {author} {\bibinfo {author} {\bibfnamefont {John~D}\ \bibnamefont
  {Joannopoulos}}, \bibinfo {author} {\bibfnamefont {Pierre~R}\ \bibnamefont
  {Villeneuve}}, \ and\ \bibinfo {author} {\bibfnamefont {Shanhui}\
  \bibnamefont {Fan}},\ }\bibfield  {title} {\enquote {\bibinfo {title}
  {Photonic crystals: putting a new twist on light},}\ }\href {\doibase
  10.1038/386143a0} {\bibfield  {journal} {\bibinfo  {journal} {Nature}\
  }\textbf {\bibinfo {volume} {386}},\ \bibinfo {pages} {143} (\bibinfo {year}
  {1997})}\BibitemShut {NoStop}%
\bibitem [{\citenamefont {Ozawa}\ \emph {et~al.}(2019)\citenamefont {Ozawa},
  \citenamefont {Price}, \citenamefont {Amo}, \citenamefont {Goldman},
  \citenamefont {Hafezi}, \citenamefont {Lu}, \citenamefont {Rechtsman},
  \citenamefont {Schuster}, \citenamefont {Simon}, \citenamefont {Zilberberg},\
  and\ \citenamefont {Carusotto}}]{Ozawa2019}%
  \BibitemOpen
  \bibfield  {author} {\bibinfo {author} {\bibfnamefont {Tomoki}\ \bibnamefont
  {Ozawa}}, \bibinfo {author} {\bibfnamefont {Hannah~M.}\ \bibnamefont
  {Price}}, \bibinfo {author} {\bibfnamefont {Alberto}\ \bibnamefont {Amo}},
  \bibinfo {author} {\bibfnamefont {Nathan}\ \bibnamefont {Goldman}}, \bibinfo
  {author} {\bibfnamefont {Mohammad}\ \bibnamefont {Hafezi}}, \bibinfo {author}
  {\bibfnamefont {Ling}\ \bibnamefont {Lu}}, \bibinfo {author} {\bibfnamefont
  {Mikael~C.}\ \bibnamefont {Rechtsman}}, \bibinfo {author} {\bibfnamefont
  {David}\ \bibnamefont {Schuster}}, \bibinfo {author} {\bibfnamefont
  {Jonathan}\ \bibnamefont {Simon}}, \bibinfo {author} {\bibfnamefont {Oded}\
  \bibnamefont {Zilberberg}}, \ and\ \bibinfo {author} {\bibfnamefont {Iacopo}\
  \bibnamefont {Carusotto}},\ }\bibfield  {title} {\enquote {\bibinfo {title}
  {Topological photonics},}\ }\href {\doibase 10.1103/RevModPhys.91.015006}
  {\bibfield  {journal} {\bibinfo  {journal} {Rev. Mod. Phys.}\ }\textbf
  {\bibinfo {volume} {91}},\ \bibinfo {pages} {015006} (\bibinfo {year}
  {2019})}\BibitemShut {NoStop}%
\bibitem [{\citenamefont {Lieb}\ \emph {et~al.}(1961)\citenamefont {Lieb},
  \citenamefont {Schultz},\ and\ \citenamefont {Mattis}}]{mattis1961}%
  \BibitemOpen
  \bibfield  {author} {\bibinfo {author} {\bibfnamefont {Elliott}\ \bibnamefont
  {Lieb}}, \bibinfo {author} {\bibfnamefont {Theodore}\ \bibnamefont
  {Schultz}}, \ and\ \bibinfo {author} {\bibfnamefont {Daniel}\ \bibnamefont
  {Mattis}},\ }\bibfield  {title} {\enquote {\bibinfo {title} {Two soluble
  models of an antiferromagnetic chain},}\ }\href {\doibase
  https://doi.org/10.1016/0003-4916(61)90115-4} {\bibfield  {journal} {\bibinfo
   {journal} {Ann. Phys. (N. Y.)}\ }\textbf {\bibinfo {volume} {16}},\ \bibinfo
  {pages} {407} (\bibinfo {year} {1961})}\BibitemShut {NoStop}%
\bibitem [{\citenamefont {Unruh}(1981)}]{Unruh:1980cg}%
  \BibitemOpen
  \bibfield  {author} {\bibinfo {author} {\bibfnamefont {W.G.}\ \bibnamefont
  {Unruh}},\ }\bibfield  {title} {\enquote {\bibinfo {title} {{Experimental
  black hole evaporation}},}\ }\href {\doibase 10.1103/PhysRevLett.46.1351}
  {\bibfield  {journal} {\bibinfo  {journal} {Phys. Rev. Lett.}\ }\textbf
  {\bibinfo {volume} {46}},\ \bibinfo {pages} {1351--1353} (\bibinfo {year}
  {1981})}\BibitemShut {NoStop}%
\bibitem [{\citenamefont {Weinfurtner}\ \emph {et~al.}(2011)\citenamefont
  {Weinfurtner}, \citenamefont {Tedford}, \citenamefont {Penrice},
  \citenamefont {Unruh},\ and\ \citenamefont {Lawrence}}]{Weinfurtner2011}%
  \BibitemOpen
  \bibfield  {author} {\bibinfo {author} {\bibfnamefont {Silke}\ \bibnamefont
  {Weinfurtner}}, \bibinfo {author} {\bibfnamefont {Edmund~W.}\ \bibnamefont
  {Tedford}}, \bibinfo {author} {\bibfnamefont {Matthew C.~J.}\ \bibnamefont
  {Penrice}}, \bibinfo {author} {\bibfnamefont {William~G.}\ \bibnamefont
  {Unruh}}, \ and\ \bibinfo {author} {\bibfnamefont {Gregory~A.}\ \bibnamefont
  {Lawrence}},\ }\bibfield  {title} {\enquote {\bibinfo {title} {Measurement of
  stimulated hawking emission in an analogue system},}\ }\href {\doibase
  10.1103/PhysRevLett.106.021302} {\bibfield  {journal} {\bibinfo  {journal}
  {Phys. Rev. Lett.}\ }\textbf {\bibinfo {volume} {106}},\ \bibinfo {pages}
  {021302} (\bibinfo {year} {2011})}\BibitemShut {NoStop}%
\bibitem [{\citenamefont {Philbin}\ \emph {et~al.}(2008)\citenamefont
  {Philbin}, \citenamefont {Kuklewicz}, \citenamefont {Robertson},
  \citenamefont {Hill}, \citenamefont {K{\"o}nig},\ and\ \citenamefont
  {Leonhardt}}]{philbin2008}%
  \BibitemOpen
  \bibfield  {author} {\bibinfo {author} {\bibfnamefont {Thomas~G}\
  \bibnamefont {Philbin}}, \bibinfo {author} {\bibfnamefont {Chris}\
  \bibnamefont {Kuklewicz}}, \bibinfo {author} {\bibfnamefont {Scott}\
  \bibnamefont {Robertson}}, \bibinfo {author} {\bibfnamefont {Stephen}\
  \bibnamefont {Hill}}, \bibinfo {author} {\bibfnamefont {Friedrich}\
  \bibnamefont {K{\"o}nig}}, \ and\ \bibinfo {author} {\bibfnamefont {Ulf}\
  \bibnamefont {Leonhardt}},\ }\bibfield  {title} {\enquote {\bibinfo {title}
  {Fiber-optical analog of the event horizon},}\ }\href {\doibase
  10.1126/science.1153625} {\bibfield  {journal} {\bibinfo  {journal}
  {Science}\ }\textbf {\bibinfo {volume} {319}},\ \bibinfo {pages} {1367}
  (\bibinfo {year} {2008})}\BibitemShut {NoStop}%
\bibitem [{\citenamefont {Steinhauer}(2016)}]{steinhauer2016}%
  \BibitemOpen
  \bibfield  {author} {\bibinfo {author} {\bibfnamefont {Jeff}\ \bibnamefont
  {Steinhauer}},\ }\bibfield  {title} {\enquote {\bibinfo {title} {Observation
  of quantum hawking radiation and its entanglement in an analogue black
  hole},}\ }\href {\doibase 10.1038/nphys3863} {\bibfield  {journal} {\bibinfo
  {journal} {Nat. Phys.}\ }\textbf {\bibinfo {volume} {12}},\ \bibinfo {pages}
  {959} (\bibinfo {year} {2016})}\BibitemShut {NoStop}%
\bibitem [{\citenamefont {Carusotto}\ \emph {et~al.}(2008)\citenamefont
  {Carusotto}, \citenamefont {Fagnocchi}, \citenamefont {Recati}, \citenamefont
  {Balbinot},\ and\ \citenamefont {Fabbri}}]{Carusotto2008}%
  \BibitemOpen
  \bibfield  {author} {\bibinfo {author} {\bibfnamefont {Iacopo}\ \bibnamefont
  {Carusotto}}, \bibinfo {author} {\bibfnamefont {Serena}\ \bibnamefont
  {Fagnocchi}}, \bibinfo {author} {\bibfnamefont {Alessio}\ \bibnamefont
  {Recati}}, \bibinfo {author} {\bibfnamefont {Roberto}\ \bibnamefont
  {Balbinot}}, \ and\ \bibinfo {author} {\bibfnamefont {Alessandro}\
  \bibnamefont {Fabbri}},\ }\bibfield  {title} {\enquote {\bibinfo {title}
  {Numerical observation of hawking radiation from acoustic black holes in
  atomic bose{\textendash}einstein condensates},}\ }\href {\doibase
  10.1088/1367-2630/10/10/103001} {\bibfield  {journal} {\bibinfo  {journal}
  {New J. Phys.}\ }\textbf {\bibinfo {volume} {10}},\ \bibinfo {pages} {103001}
  (\bibinfo {year} {2008})}\BibitemShut {NoStop}%
\bibitem [{\citenamefont {Koll{\'a}r}\ \emph {et~al.}(2019)\citenamefont
  {Koll{\'a}r}, \citenamefont {Fitzpatrick},\ and\ \citenamefont
  {Houck}}]{kollar2019hyperbolic}%
  \BibitemOpen
  \bibfield  {author} {\bibinfo {author} {\bibfnamefont {Alicia~J}\
  \bibnamefont {Koll{\'a}r}}, \bibinfo {author} {\bibfnamefont {Mattias}\
  \bibnamefont {Fitzpatrick}}, \ and\ \bibinfo {author} {\bibfnamefont
  {Andrew~A}\ \bibnamefont {Houck}},\ }\bibfield  {title} {\enquote {\bibinfo
  {title} {Hyperbolic lattices in circuit quantum electrodynamics},}\ }\href
  {\doibase 10.1038/s41586-019-1348-3} {\bibfield  {journal} {\bibinfo
  {journal} {Nature}\ }\textbf {\bibinfo {volume} {571}},\ \bibinfo {pages}
  {45} (\bibinfo {year} {2019})}\BibitemShut {NoStop}%
\bibitem [{\citenamefont {Boettcher}\ \emph {et~al.}(2020)\citenamefont
  {Boettcher}, \citenamefont {Bienias}, \citenamefont {Belyansky},
  \citenamefont {Koll\'ar},\ and\ \citenamefont {Gorshkov}}]{Boettcher2020}%
  \BibitemOpen
  \bibfield  {author} {\bibinfo {author} {\bibfnamefont {Igor}\ \bibnamefont
  {Boettcher}}, \bibinfo {author} {\bibfnamefont {Przemyslaw}\ \bibnamefont
  {Bienias}}, \bibinfo {author} {\bibfnamefont {Ron}\ \bibnamefont
  {Belyansky}}, \bibinfo {author} {\bibfnamefont {Alicia~J.}\ \bibnamefont
  {Koll\'ar}}, \ and\ \bibinfo {author} {\bibfnamefont {Alexey~V.}\
  \bibnamefont {Gorshkov}},\ }\bibfield  {title} {\enquote {\bibinfo {title}
  {Quantum simulation of hyperbolic space with circuit quantum electrodynamics:
  From graphs to geometry},}\ }\href {\doibase 10.1103/PhysRevA.102.032208}
  {\bibfield  {journal} {\bibinfo  {journal} {Phys. Rev. A}\ }\textbf {\bibinfo
  {volume} {102}},\ \bibinfo {pages} {032208} (\bibinfo {year}
  {2020})}\BibitemShut {NoStop}%
\end{thebibliography}%

\end{document}